# Recent neutron focusing experiments using polycapillary lens in CSNS[*]

Kai Pan(潘凯)[1,2,3,&], Xue-Peng Sun(孙学鹏)[1,3,&], Tian-Cheng Yi(易天成)[4,5], Song-Ling Wang(王松林)[4,5], Jian-Rong Zhou(周健荣)[4,5], Mo Zhou(周末)[3], Xing-Fen Jiang(蒋兴奋)[4,5], Bin Zhou(周斌)[4,5], Bo-Wen Jiang(姜博文)[6], Tian-Xi Sun(孙天希)[1,2,3], Tian-Jiao Liang(梁天骄)[4,5,†] and Zhi-Guo Liu(刘志国)[1,2,3,†]

[1] *College of Nuclear Science and Technology, Beijing Normal University, Beijing 100875, China*
[2] *Beijing Key Laboratory of Applied Optics, Beijing 100875, China*
[3] *Key Laboratory of Beam Technology, Ministry of Education, Beijing 100875, China*
[4] *Spallation Neutron Source Science Center, Dongguan 523803, China*
[5] *Institute of High Energy Physics, Chinese Academy of Sciences, Beijing 100049, China*
[6] *North Night Vision Technology (Nanjing) Research Institute Co.,Ltd, Nanjing 211106, China*

Higher neutron flux densities can provide convenience for neutron experiments. Using neutron optical focusing elements, large flux beams transported to sample can be achieved. As one kind of focusing elements, polycapillary lens is very suitable for neutron activation experiments such as PGAA and NDP technology. Such experiments in China Spallation Neutron Source (CSNS) are interested areas to researchers. To provide some suggestions and ideas for the following design of enhanced PGAA or NDP instrument with polycapillary lens in CSNS, a first neutron focusing experiment using polycapillary lens in CSNS was conducted. For 1.0-12.6 Å polychromatic beam, a focal spot with FWHM of 800 μm was obtained. In detailed wavelength-dependent measurements, the spot size, transmission efficiency and gain increased as the value of wavelength increased. For cold neutrons, the gain maintained in a level of 7.

## 1. Introduction

As neutron microanalysis technique is changing from one-dimensional to two-dimensional, there is a demand trend for sufficient measurements. This can be achieved with large flux beams transported to sample. Actually, available neutron sources usually have a lower brilliance compared with X-ray sources provided by synchrotron radiation which limits the flux value. Higher neutron flux densities can provide convenience for neutron experiments, especially neutron activation experiments such as NDP (Neutron Depth Profile) and PGAA (Prompt Gamma Activation Analysis). It can be realized by using neutron optical focusing element. These neutron optical elements are mainly based on three different kinds of manipulation principles, namely diffraction, refraction and reflection [1] In the past several decades, the field of reflective optics is the largest field for neutron focusing. Plenty of solutions such as Wolter mirrors [2], "lobster eye" lenses [3], polycapillary lenses [4] and so on were proposed. Reflective neutron optics are free from chromatic aberration which limits the effectiveness of refractive and diffractive neutron optics when used with a broad wavelength neutron beam.

---

[*] Project supported by the National Natural Science Foundation of China (Grant No. 12175021), the National Natural Science Foundation of China (Grant No. 12175254) and the Guangdong Basic and Applied Basic Research Foundation (Grant No. 2019A1515110398).
[†] Corresponding author. E-mail: tjliang@ihep.ac.cn, liuzhiguo512@126.com
[&] Both authors contributed equally to this work.

Polycapillary lenses is a standard manipulation optics for X-ray beams. It is also available for neutrons. Internationally, polycapillary lenses have been used for neutron focusing and related results were reported. An early polycapillary neutron lens was proposed by M. A. Kumakhov and V. A. Sharov [5], and it amplified the neutron flux density by a factor of 6-7 at the Kurchatov Institue IR-8 reactor. Subsequent works were mainly completed by D. F. R. Mildner and H. H. Chen-Mayer at the NIST Research Reactor and on the Intense Pulsed Neutron Source (IPNS) [6-8]. In these works, large entrance area lead-glass or boron glass monolithic lenses were used. They gave a focal spot of size ~100 μm and the gain were in the range of 10 to 100. This component is very efficient when implemented for PGAA and NDP applications. In 1995, placing a focusing neutron lens into a PGAA instrument indicated that this technique increased the elemental sensitivities of PGAA [9]. In 1997, polycapillary lens was used to focus a neutron beam to sub-millimeter spot size [10]. This technique adds lateral spatial sensitivity to NDP and makes compositional mapping in three dimensions possible. In addition, using a monolithic polycapillary focusing lens for neutron diffraction application can overcome the limitation of unavailable large crystals [11].

China Spallation Neutron Source (CSNS) [12], as the first pulsed neutron source in China, is a large scientific platform. It is constructed to investigate neutron properties and to study the microstructure and motion of matter with the help of corresponding neutron detection techniques. At present, fewer literatures about the researches of the use of neutron optical elements in CSNS have reported. Though, using a polycapillary focusing lens will bring considerable benefits for PGAA and NDP experiments as mentioned above, there existed a blank in these kinds of experiments which implemented in CSNS. This paper reports the first focusing experiment using monolithic polycapillary focusing lens in CSNS. The whole experiment includes several measurements which taken to determine the transmission characteristics of the lens. Simulation calculations of the optic transmission, spot size and gain were compared with experimental results. These results indicate that the parameters of the neutron polycapillary should be optimized for a given instrument with its particular spectral characteristics. This work will provide some suggestions and ideas for the following design of enhanced PGAA or NDP instrument with polycapillary lens in CSNS.

## 2. Lens and experimental arrangement
### 2.1. Polycapillary neutron focusing lens

The monolithic polycapillary neutron focusing lens has a hexagon cross-section and consists of a bundle of curved fibers. An individual fiber is fabricated by thousands of hollow glass channels of diameter in micron dimension. The whole structure can be simplified into two parts: the parallel part and the curved part. The channels are parallel at the entrance and used to collect the radiation emitted from the source. Then multiple glancing reflections of the radiation occur at the air-glass boundary of the curved part and finally deflected into a focused beam. In short, a parallel beam with divergence can be converged to a small spot based on the principle of total external reflection. The focal spot is the result of superposition of the many small beams from the hollow channels. The detailed parameters of the polycapillary focusing lens used in this work was summarized in Table. 1. Fig. 1 showed the photography of some lens fabricated by Beijing Normal University (BNU). The shortest one, also called the semi-lens, was used as the focusing lens in this work.

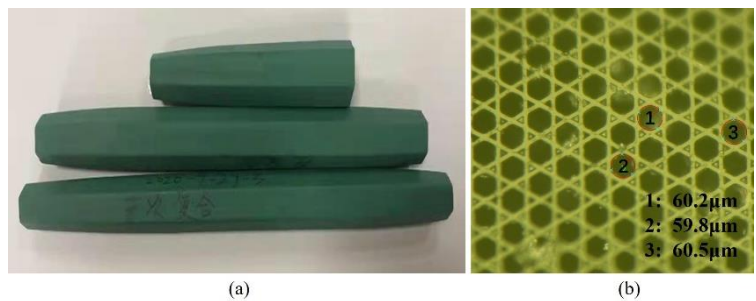

(a)          (b)

Fig.1 (a)The photography of neutron polycapillary lenses fabricated by BNU, the focusing lens (the shortest) was used in this work; (b)Electron micrography of the cross-section of the shortest lens.

Table. 1 Detailed parameters of the polycapillary neutron focusing lens

| Geometric structure | | | | |
|---|---|---|---|---|
| Entrance(mm) | Exit(mm) | Length(mm) | Open area ratio | Channel diameter(μm) |
| 20.0 | 15.0 | 60.0 | 0.75 | 60.0 |
| Chemical composition of glass(wt%) | | | | | | |
| $SiO_2$ | $B_2O_3$ | $K_2O$ | $Al_2O_3$ | $Na_2O$ | CaO | $Fe_2O_3$ |
| 75.50 | 15.00 | 0.60 | 3.40 | 4.70 | 0.40 | 0.05 |

### 2.2. Experimental arrangement

The whole focusing experiment for the polycapillary lens was conducted on the beamline 20 (BL20) in CSNS. BL20 of CSNS was operated as a specialized neutron test beamline for neutron beam quality analysis and spectrometer detector test. It provides polychromatic neutron beam of 1.0-12.6 Å wavelength bandwidth. The flux of BL20 is about $10^6 \, n/cm^2/s$. The angular divergence of the neutron beam is expected to be $\pm 6 \, mrad$ which is measured using imaging plates at the different distance away from the aperture 2.

Fig. 2 shows the schematic diagrams of experimental set-ups and arrangement. Aperture 1 is the exit of the neutron beam of BL20. A 20 mm aperture 2 was attached in the back-end to confirm that the input beam size matches the entrance area of the lens in rough. The lens was fixed on the sample stage with five degrees of freedom using an aluminous holder. For the remotely controlled sample stage, the three orthogonal translations were with 150 mm range and 2 μm precision; the two orthogonal angular motions were with $0.003°$ and $0.006°$ precision. The distance between the exit of the aperture and entrance of the lens was 1.5 m. The alignment and measurements of the lens were conducted using two different detector system in this work. Firstly, a computer-controlled CCD (charge coupled device) camera was used to observe the neutron beam in broadband wavelength, the pixel size is 15 μm, the pixel number is $2048 \times 2048$, the magnification factor of the camera lens is 0.575. As shown in Fig. 2(a), the T0 chopper can stop the gamma rays and low wavelength neutrons (less than 1 Å) and the T1 chopper can provide fixed bandwidth of 4.5 Å neutrons. Secondly, a TPX3Cam camera was used to characterize the focusing performance of the lens as a function of wavelength. This TPX3Cam camera is an energy resolved neutron imaging detector with high timing resolution and spatial resolution. The pixel size of this camera is 55 μm, the pixel number is $256 \times 256$, the magnification factor of the camera lens is 1.000. As shown in Fig. 2(b), the T0 signal represents the time of the generation of the pulsed neutron beam. The TPX3Cam of the camera records the coordinates (x, y), the time of arrival (TOA) and time over threshold (TOT) for the pixels fired almost at the same time. The represented energy or wavelength of the neutron that detected by the camera was calculated according to T0 and TOA. The actual experimental arrangements for the measurements was shown in Fig .3.

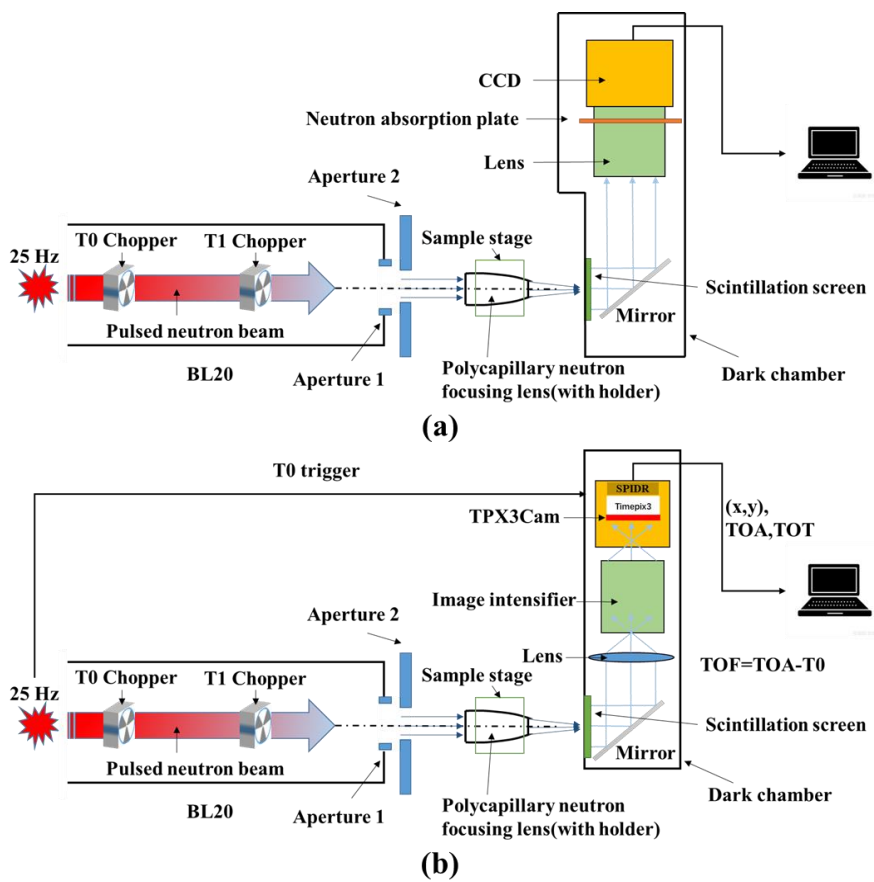

Fig. 2 (a)Schematic diagram of experimental set-ups (with CCD camera); (b) Schematic diagram of experimental set-ups (with TPX3Cam camera).

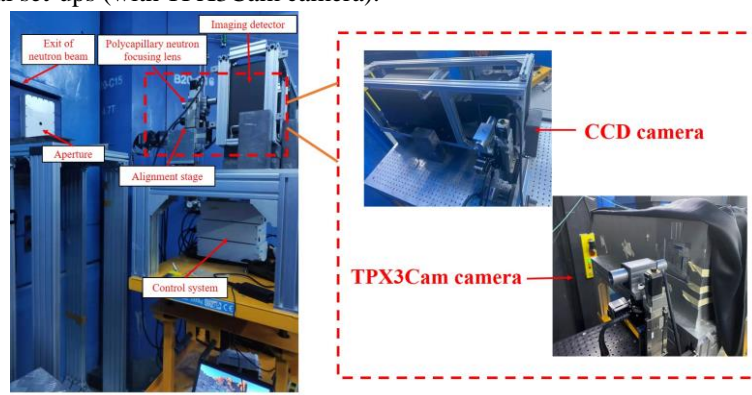

Fig. 3 The actual experimental arrangement for the measurements

## 3. Computer simulation

In this work, the design of a polycapillary neutron focusing lens was considered for focusing thermal and cold neutrons. D.F.R. Milder and H. Chen developed the criteria for characterizing neutron focusing lens. [13] The derived expressions can be used for initial determination of the parameters of the lens. For a more precise value, a Monte-Carlo simulation program was needed. We used C++ language to code such a program for calculating the trajectories of a large number of neutrons that accepted by the lens. The detailed model was based on our previous works on X-ray polycapillary lens and the physic process considered was referred to reference [14], [15]. In detail, the simulation mainly includes three parts: source sampling, neutron beam tracing and data statistics. Source sampling takes account of neutron energy and position. The position of neutron emitted from the source was sampled using Gibbs sampling algorithm. Fig. 4(a) shows the intensity distribution of the neutron beam on the exit plane of the aperture that recorded by the CCD camera in 10 seconds. Fig. 4(b) to (d) shows the results of position sampling for different sampling event numbers. For a specified position, corresponding intensity information was deemed to be confirmed. After position sampling, sampling of wavelength was conducted successively using the Metropolis-Hastings sampling algorithm. In the process of wavelength sampling, the spectra of the source were normalized. Fig. 5 shows the results of wavelength sampling.

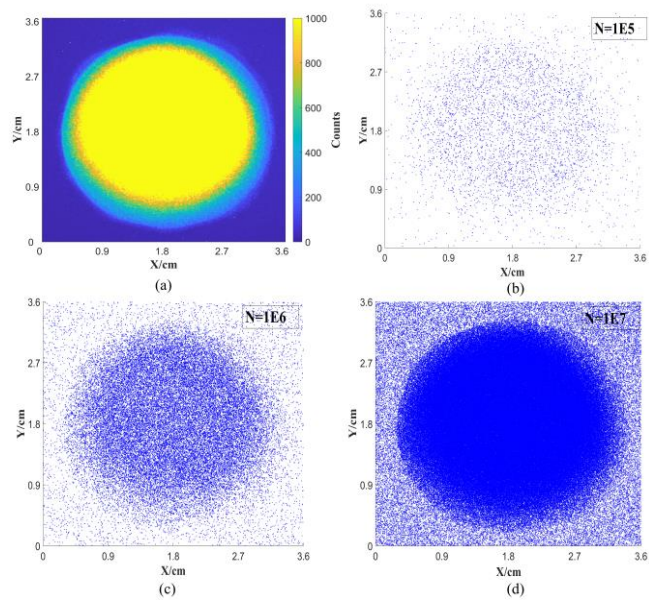

Fig. 4 (a) The intensity distribution of the neutron beam on the exit plane of the aperture that recorded by the CCD camera in 10 seconds; (b)-(d) Results of position sampling for different sampling event numbers: $10^5$, $10^6$ and $10^7$.

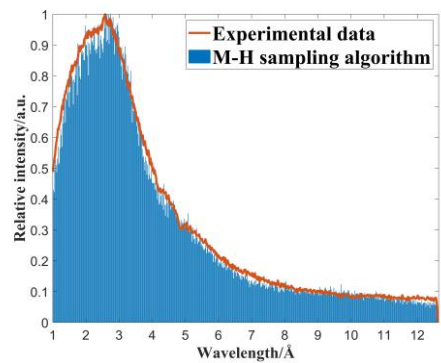

Fig. 5 Wavelength sampling results, the sampling interval is 0.2 Å and the sampling event number is $10^6$.

When the position, relative intensity and wavelength of a neutron were confirmed through the source sampling process, its tracing process was started. The initial energy of the neutron can be calculated using the following formula:

$$E(eV) = \frac{1}{2m_n}\left(\frac{h}{\lambda}\right)^2 = \frac{1}{2m_n c^2}\left(\frac{hc}{\lambda}\right)^2 = \frac{0.082}{\lambda^2(\text{Å})} \tag{1}$$

Where $h$ is Plank constant, $m_n$ is residual neutron mass and $\lambda$ is neutron wavelength.

The intensity value of a neutron was also normalized to 1 according to the intensity map recorded by the CCD camera for its first collision in air-glass boundary. Then the probability that the neutron is actually transmitted was obtained as the product of the reflection probabilities for each reflection. The reflection probability is calculated using Fresnel equation and the theoretical neutron reflection coefficient as a function of incident angle is shown in Fig .6. In addition, the neutron attenuation in the glass was taken into account.

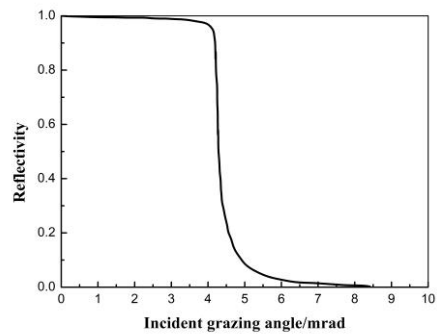

Fig. 6 The theoretical neutron reflection coefficient as a function of incident angle

Finally, the neutron trajectories intersect the detector plane, its position and intensity were summed on a uniform grid. Fig. 7 shows the simulation results of the intensity distribution of the neutron beam when the detector plane was placed in different positions longitudinally. The simulation results indicate that with the distance between the lens exit and detector plane increases, the neutron beam was converged gradually first and then diverged. The FWHM of the focal spot is 700 μm and the focal length is 60.0 mm.

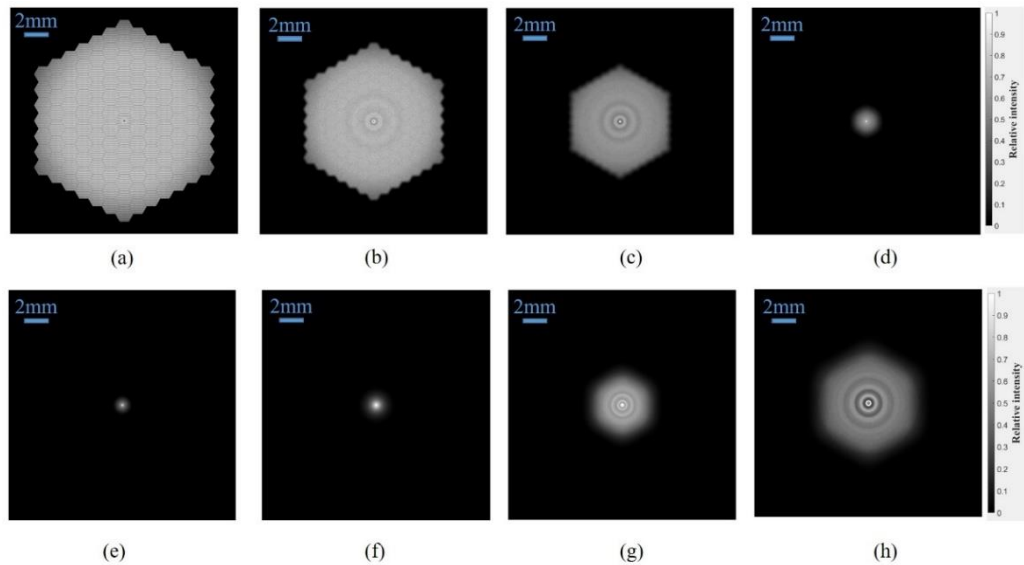

Fig. 7 Transverse intensity distribution of the neutron beam for different distance between the lens exit and detector plane: (a) 0.0 mm; (b) 15.0 mm; (c) 30.0 mm; (d) 45.0 mm; (e) 60.0 mm; (f) 75.0 mm; (g) 90.0 mm; (h) 105.0 mm.

4. **Results and discussion**

The neutron beam at the optic focus was observed by the CCD camera after correct alignment. Fig. 8(a) shows an image of the focal spot with polychromatic neutrons (1.0-12.6 Å) when the camera was placed 61.0mm away from the lens exit, and the exposure time was set to 60 seconds. The image of Fig. 8(a) was processed using the histogram equalization for better distinction. In the center of the image, a very intense spot of light exists. Neutrons focused by the polycapillary focusing lens produced the spot. The hexagon defines the edges of the lens and it constitutes main part of the background. The bright ring of the hexagon is due to transmission light marked in the figure. These light shows that the neutrons, especially for thermal neutrons of short wavelength are not captured and guided by the lens. They penetrate through the glass walls of the channels with absorption and finally projected on the detector plane. Another kind of transmission light is due to neutrons that penetrate through the aluminum with attenuation loss. The two circular light spots are caused by neutrons transmit directly through the holes which provides convenience for insertion of the boron carbide annulus into the lens holder. The annulus was used for lens support and radiation shield of the neutrons in the gaps. Fig. 8(b) is the raw image and (c) is the intensity distribution of the spot in one dimension. The FWHM of the focal spot is about 800 μm.

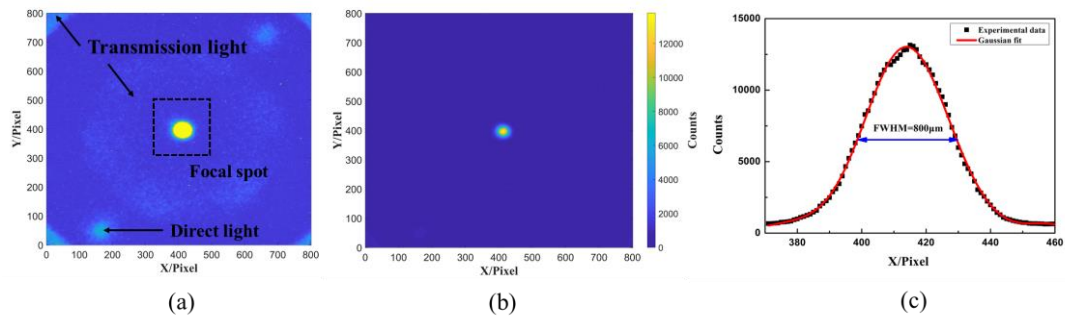

Fig. 8 (a)Image of the focal spot with polychromatic neutrons processed; (b)Raw image of the focusing performance; (c)Intensity distribution of the focal spot in one dimension.

From Fig. 8, it clearly indicates that the polycapillary lens is suitable for polychromatic neutron beam in CSNS. For a broadband wavelength, it can produce a hundred micron-scaled spot. The focal spot size is larger than simulation results for the imperfect superposition of small beams from each individual channel. To investigate the convergence of the neutron beam, depth-dependent measurements in a scan along the longitude direction were conducted. Fig. 8 shows the measured FWHM for a longitudinal scan for 1.0-12.6 Å polychromatic beam. Taking the minimum value of beam size as the zero, the focal depth measured was about 20mm in Fig. 9(a). The half-convergence and half-divergence angle was $0.78°$ and $0.83°$. In Mildner's theorem, the calculated half-convergence angle is $\Omega = R_E / L_F = 7.5 \times 180° / (60 \times \pi) = 7.16°$. Here, $R_E$ is the radius of the lens exit and $L_F$ is the focal length. The difference between the calculated and measured results demonstrate the low transmission efficiency of the outer and more curved channels. In Fig. 9(b), the experimental results in vertical and horizontal are similar which is the result of the symmetry of the fabricated lens. The Monte-Carlo result is similar to the theoretical, the FWHM of the focal spot size is 700 μm and the half-convergence angle is $7°$.

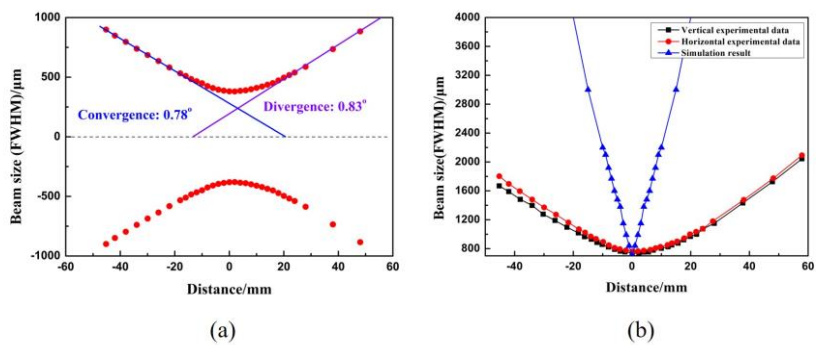

Fig. 9 (a)Measured FWHM for a longitudinal scan; (b)Experimental results in vertical and horizontal versus simulation result

Fig. 10 shows a summary of all the measurements taken for different wavelength over a fixed bandwidth. The starting wavelength is 1 to 6 Å and the fixed bandwidth is 5 Å. The polychromatic neutron beam with fixed bandwidth was realized using a neutron chopper. Here, the transmission efficiency of the lens is defined as the ratio of the detector counts of output neutrons at the exit of the lens to that of the input neutrons at the entrance of the lens. The gain of the lens is determined by the ratio of the detector counts within the FWHM for a Gaussian distribution fitted to the measured peak to that within a spot of the same size without the lens. In total, the gain, transmission efficiency and FWHM of the focal spot increases with increasing incident wavelength. These are the results of higher critical angle for lower incident energy.

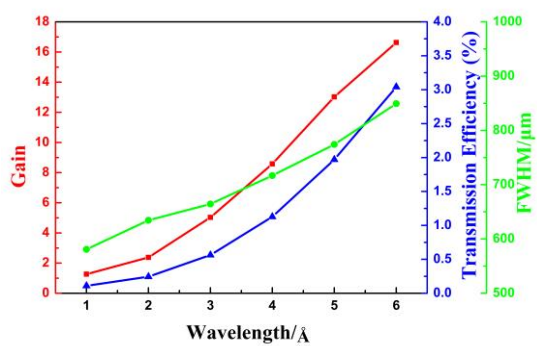

Fig. 10 Measured gain, transmission efficiency and FWHM of the focal spot with increasing incident wavelength

For detailed investigation of the transmission properties of the lens as a function of wavelength, the TPX3Cam camera was used. Fig. 11 shows the focal spot images recorded by the camera when the wavelength of the TOF mode was set to 1 Å. Intuitively, the lens started to work well when the incident wavelength is above 3 Å. Fig. 12 shows the longitudinal scan results. The gain was obtained by dividing the intensity of the direct beam (without the focusing lens) into the focused beam. Over the bandwidth from 2.7 to 12.6 Å, the gain values are larger than 1 in the range of focal depth. This indicates that for incident neutrons with wavelength shorter than 2.7 Å, the polycapillary lens cannot work well. These neutrons penetrate the glass with attenuation loss and constitute the background.

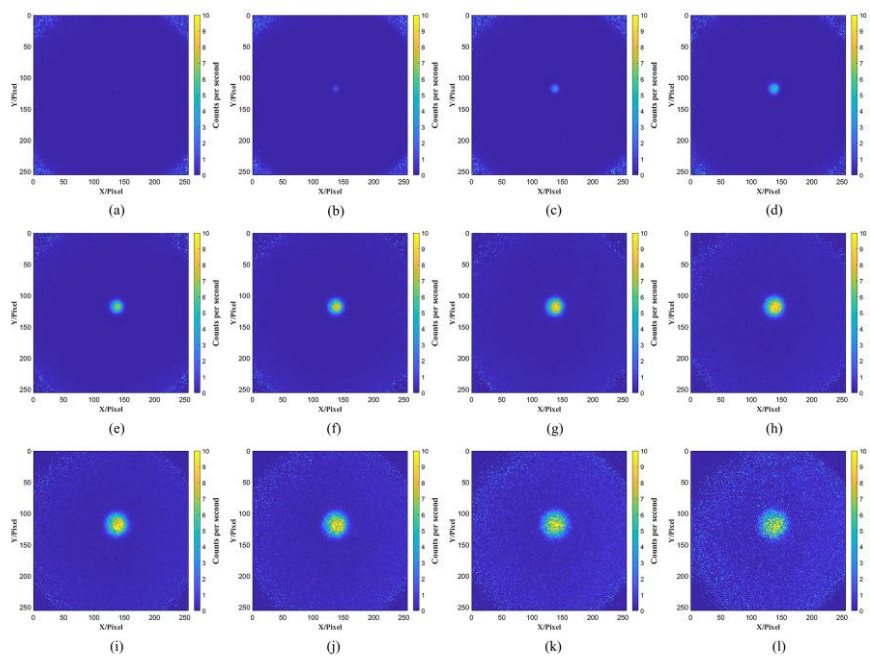

Fig. 11 Focal spot images for different incident neutron wavelength: (a)1-2 Å ; (b)2-3 Å ; (c)3-4 Å ; (d)4-5 Å ; (e)5-6 Å ; (f)6-7 Å ; (g)7-8 Å ; (h)8-9 Å ; (i)9-10 Å ; (j)10-11 Å ; (k)11-12 Å ; (l)12-12.6 Å .

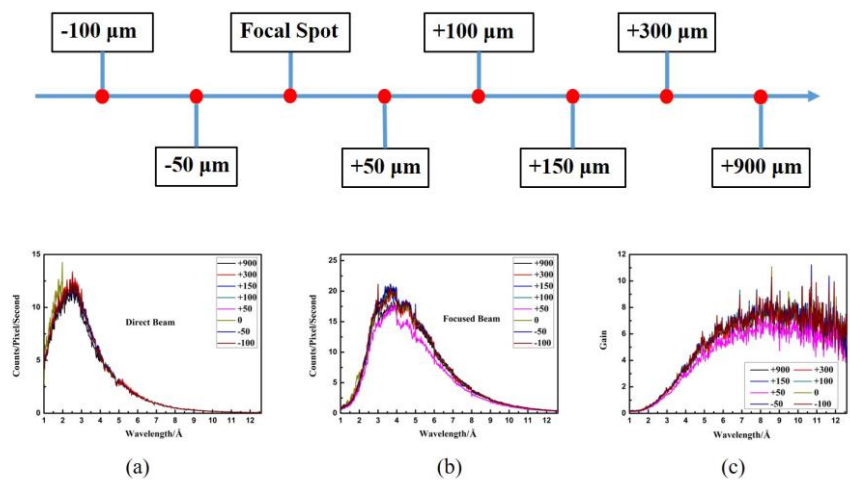

Fig .12 Measurement results for longitudinal scan: (a)direct beam; (b)focused beam; (c)gain

Fig. 13 summarized all of the measurements of the beam size and gain. Corresponding calculated and simulation results are given. All of the results in Fig. 13(a) shows that the beam size increased with increasing wavelength. The spots observed below 4 Å are small when the experimental data was handled with background correction while the calculated and simulation results are similar and bigger. The gain increases below 6 Å  and then decreases. The peak value is related to the divergence angle of incident beam and the value of the critical angle. When the incident wavelength is 1 Å , the gain value is less than 1 which means the spot is due to neutron penetration in central channels. For wavelength between 1 to 3 Å , the values of calculated and simulation results of the beam size are larger than experimental results. When the wavelength is over 4 Å , the relationship is reversed. In Fig. 13(b), the calculated and simulation results of gain are larger than experimental results in the whole band. This can be explained by imprecise fabricated outline of the lens which caused the low transmission efficiency of the outer channels. In the meanwhile, we can conclude that the focusing performance of the lens in short wavelength is influenced effectively by penetration.

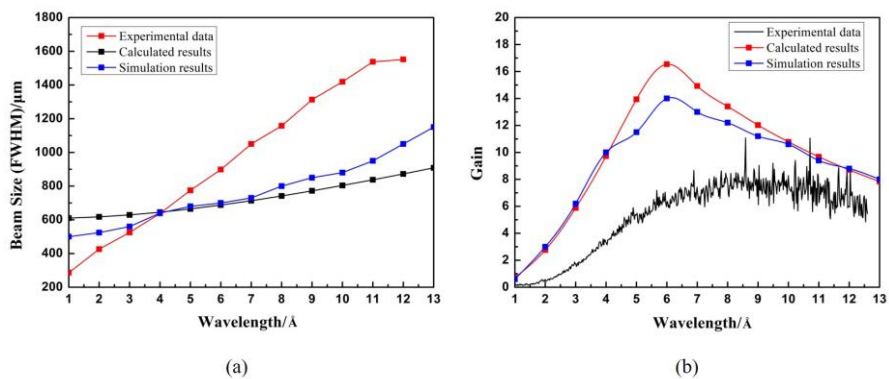

Fig. 13 (a)Beam size with increased incident wavelength; (b)Gain with increased incident wavelength

## 5. Conclusion

Using a polycapillary focusing lens in NDP and PGAA experiments will bring considerable benefits. The hundred micron-scaled spot obtained using this kind of lens makes 3D NDP possible and improves the spatial resolution of NDP to a sub-millimeter level. It can also increase the elemental sensitivities of PGAA. In this work, we presented our measurement results of the performance of the neutron polycapillary focusing lens in CSNS. The aim of these measurements is to provide some suggestions and ideas for design of enhanced PGAA or NDP instrument with polycapillary lens in CSNS.

In the design process of the neutron lens, a program including source sampling, neutron beam tracing and data statistics was coded and used for simulation. Expressions derived by Mildner and Chen also used for comparison. The following experiments was conducted in BL20 of CSNS using a CCD camera and a TPX3Cam camera with TOF mode. Placing the CCD camera at the focus, a very intense spot of light with the FWHM of $800\,\mu m$ was observed. For a longitudinal scan, the measured half-convergence angle was much smaller than the calculated and simulated results. This is because of the low transmission efficiency of the outer channels of the lens due to the imprecise fabrication. In rough, a wave chopper was used to select a fixed bandwidth with different starting wavelength value. The gain, transmission efficiency and FWHM of the focal spot increases with increasing incident wavelength. For detailed investigation, a TPX3Cam camera was used to repeat those measurements. Intuitively, the lens started to work when the incident wavelength is above $3\,\text{Å}$ according to the spot images. From the longitudinal scan results, it demonstrated that the lens works well over the bandwidth from 2.7 to $12.6\,\text{Å}$. The focusing performance of the lens in short wavelength is influenced effectively by penetration. Compared to calculated and simulated results, the experimental results at present are low.

Compared to the international level, the gain of the lens is still low while it has been verified that the lens is an effective neutron optical element. Higher transmission efficiency and gain of the neutron polycapillary focusing lens are the next goal of our research work. Construction of enhanced PGAA and NDP instruments with polycapillary lens in CSNS is our near-term expectation.

**Acknowledgment**

This work was supported by the National Natural Science Foundation of China (Grant No. 12175021), the National Natural Science Foundation of China (Grant No. 12175254) and the Guangdong Basic and Applied Basic Research Foundation (Grant No. 2019A1515110398). The authors thank Ming-Nian Yuan and Wen-Qin Yang for their hard work in the whole process of fabrication of the lens and the performance measurement, respectively.